\documentstyle[prb,aps,epsf]{revtex}

\begin{document}
\draft

\twocolumn[\hsize\textwidth\columnwidth\hsize\csname @twocolumnfalse\endcsname

\title
{\bf Diffusion of Pt dimers on Pt(111)}

\author{Ghyslain Boisvert\cite{byline1} and Laurent J. Lewis\cite{byline2}}

\address{
D{\'e}partement de Physique et Groupe de Recherche en Physique et Technologie des
Couches Minces (GCM), Universit{\'e} de Montr{\'e}al, Case Postale 6128, Succursale
Centre-Ville, Montr{\'e}al, Qu{\'e}bec, Canada H3C 3J7
}

\date{\today}

\maketitle

\begin{center}
{Submitted to Physical Review B}
\end{center}

\begin{abstract}

We report the results of a density-functional study of the diffusion of Pt
dimers on the (111) surface of Pt. The calculated activation energy of 0.37
eV is in {\em exact} agreement with the recent experiment of Kyuno {\em et
al.} \protect{[}Surf. Sci. {\bf 397}, 191 (1998)\protect{]}. Our calculations
establish that the dimers are mobile at temperatures of interest for adatom
diffusion, and thus contribute to mass transport. They also indicate that the
diffusion path for dimers consists of a sequence of one-atom and (concerted)
two-atom jumps.

\end{abstract}


\vskip2pc
]

\narrowtext

\section{Introduction}\label{intro}

The diffusion of adatoms --- or very small clusters of them --- on clean,
infinite, defect-free surfaces remains, in spite of its conceptual
simplicity, a largely unresolved problem. Experimentally, direct measurements
of the trajectories of individual particles are possible (using field-ion
microscopy) for only a few elements (notably Ir and Pt)\cite{kellogg} in a
narrow range of temperatures. In most cases, the diffusivities are determined
indirectly, e.g., by inferring them from island growth measurements (see for
instance Ref.\ \onlinecite{venables}). The situation is just as difficult
theoretically: though the diffusion constants can in principle be determined
explicitly using, e.g., molecular-dynamics simulations, this is in practice
extremely difficult because of computer and model-potential limitations.

The (111) surface of Pt is of particular interest. The diffusivity of adatoms
has been measured using different approaches, and an activation barrier of
0.25-0.26 eV has been determined.\cite{bott,feibelman1,kyuno} This is in
disagreement with our first-principles result of 0.33 eV,\cite{boisvert98}
but within the error bar of a recent calculation by
Feibelman,\cite{feibelman2} which gives a barrier of 0.29 eV.\cite{bogicevic}
Though the agreement between theory and experiment can be judged
satisfactory, questions remain concerning the role played by dimers in the
kinetics of growth, and in particular the shape of islands as a function of
temperature.

We have carried out a series of density-functional-theory calculations in
order to determine the barrier and mechanism for diffusion of the Pt dimer on
Pt(111). The activation energy we obtain --- 0.37 eV --- is in striking (and
somewhat surprising) {\em exact} agreement with experiment.\cite{kyuno} Our
calculations therefore indicate that the dimers are certainly mobile at
temperatures of interest for adatom diffusion (and thus contribute to mass
transport), as in fact can also be inferred from the measurements of Kyuno
{\em et al.}\cite{kyuno} Our study reveals that the diffusion path for dimers
consists of a sequence of one-atom and (concerted) two-atom jumps.

\section{Computational details}\label{model}

As already noted above, the calculations reported here were performed within
the framework of density-functional theory.\cite{kohn} Our previous studies
of this surface\cite{boisvert98} indicate that the local-density
approximation (LDA)\cite{ceperley} to the exchange-and-correlation energy
provides a better description of platinum than the generalized-gradient
approximation (GGA);\cite{perdew} the LDA was therefore used for all
calculations reported here. The ion cores were approximated by
pseudopotentials with $5d$ electrons treated as valence states. The
pseudopotentials were generated using the semi-relativistic scheme of
Troullier and Martins\cite{troullier} and expressed in the Kleinman-Bylander
form using the $s$ component as the local one.\cite{kleinman,gonze,fuchs} The
electronic wave-functions were represented using a plane-wave basis set with
kinetic energy up to 40 Ry. In order to improve convergence, the electronic
states were occupied according to a Fermi distribution with $k_BT_{\rm el} =
0.1$ eV and the total energies obtained by extrapolating to zero electronic
temperature. For similar reasons, the calculations were initiated using
wave-functions obtained from the self-consistent solution of the Kohn-Sham
Hamiltonian in a mixed basis set composed of pseudo-atomic orbitals and plane
waves cut off at 4 Ry.\cite{kley} The minimization of the energy with respect
to the electronic degrees of freedom was done using an iterative
procedure.\cite{stumpf} After achieving electronic convergence, the atoms
were moved according to a damped Newton dynamics until forces became less
than $0.01$ eV/\AA. In view of the high energy cutoff needed in the
plane-wave expansion, it is important to keep the system size to a minimum.
To do so, we used a slab geometry consisting of four $3\times3$ layers (plus
vacuum in the supercell approach) with the dimer adsorbed on one side. Only
the top layer (plus the dimer) was allowed to relax. The {\bf k}-space
integration was performed using a $2\times2$ grid, the exact number of points
depending on the actual symmetry of the configuration under consideration.

\section{Results}\label{res}

\subsection{Binding and dissociation energies}\label{binding}

The binding energy of the dimer is given by $E_{\rm binding} = E_{\rm dimer}
+ E_{\rm clean} - 2 E_{\rm adsorption}$, where $E_{\rm dimer}$ is the total
energy of the system including the dimer, $E_{\rm clean}$ is the total energy
of the system with a ``clean'' surface, and $E_{\rm adsorption}$ is the
adsorption energy of a single adatom.

A single adatom on the (111) surface can sit either in a fcc site or in a
(stacking-fault) hcp-like site. The latter lies 0.17-0.21 eV above the former
in the case of platinum\cite{boisvert98,feibelman2} i.e., is very
unfavourable. For the equilibrium state of the dimer, one therefore expects
the two atoms to sit in nearest-neighbour fcc sites, such as the f$_1$f$_2$
configuration in Fig.\ \ref{surf}. The corresponding configuration where the
two atoms are in hcp-like sites, such as h$_1$h$_2$, indeed lies 0.30 eV
above f$_1$f$_2$, we have verified. In both the f$_1$f$_2$ and h$_1$h$_2$
configurations, the dimer bond length is close to $a/\sqrt{2}$ (with $a$ the
lattice parameter), the nearest-neighbour distance on the (111) surface.
Other possible configurations of the dimer have longer or shorter bond
lengths and are therefore unlikely. We have examined the f$_1$f$_3$
configuration, where the dimer atoms are second nearest neighbours (bond
length $= \sqrt{3}a/\sqrt{2}$) and found it to lie 0.56 eV above f$_1$f$_2$.
Likewise, the possible fcp-hcp configurations (such as f$_1$h$_1$,
f$_1$h$_2$, and f$_1$h$_3$ --- cf.\ Fig.\ \ref{surf}) all lie substantially
higher in energy than f$_1$f$_2$, as we will see below.

With the dimer in the f$_1$f$_2$ configuration, we obtain a binding energy of
0.48 eV, with an error that we estimate to be of the order of 0.05 eV. This
is much larger than the diffusion barrier for a single adatom, 0.33 eV.
\cite{boisvert98} The barrier towards dissociation is given, roughly, by the
sum of the dimer binding energy and the diffusion barrier of the
adatom\cite{bartelt1} --- 0.81 eV in the present case. We therefore expect
the dimers to be stable, i.e., unlikely to dissociate, in the temperature
range in which the adatoms are mobile. More precisely, the first step towards
dissociation is most likely related to the f$_1$f$_2 \rightarrow$ f$_1$h$_3$
(rather than f$_1$h$_2$) transition (from geometrical considerations: the
f$_1$h$_3$ distance is larger than the f$_1$h$_2$ one; also, there are two
recombination paths for f$_1$h$_2$, but only one for f$_1$h$_3$). We find the
barrier for this process to be about 0.75 eV, remarkably close to the above
estimate.

\subsection{Diffusion by successive jumps}\label{success}

\begin{figure}
\vspace*{-0.25in}
\epsfxsize=3.5in \epsfbox{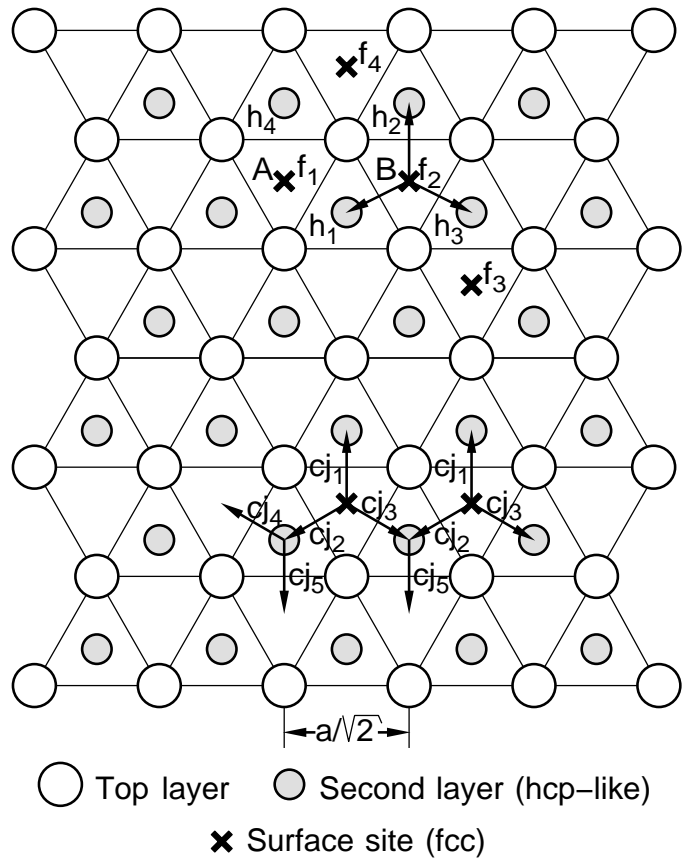}
\vspace{0.15in}
\caption{
The dimer diffusion processes studied in the present work; the symbols are
explained in the text.
\label{surf}
}
\end{figure}

We consider first the possibility that diffusion of the dimer proceeds by
successive jumps of its constituent atoms. Referring to Fig.\ \ref{surf}
again, we will assume that atom B, on the f$_2$ site, jumps first, followed
by atom A, on the f$_1$ site. B can jump to either of the three adjacent
hcp-like sites, as indicated by arrows in Fig.\ \ref{surf}. The f$_1$h$_i$
($i=$ 1, 2, 3) configurations of the dimer are metastable states since hcp
sites are not equilibrium sites on this surface. Evidently, these states will
be relevant to diffusion only if they lie sufficiently low in energy above
the equilibrium state --- more specifically by an amount which is of the
order of the barrier for single-atom diffusion (0.33 eV).

We have calculated the energies of the f$_1$h$_i$ configurations of Fig.\
\ref{surf}, and obtain, as measured with respect to the f$_1$f$_2$ state,
$\Delta E = $ 0.87, 0.34, and 0.69 eV for f$_1$h$_1$, f$_1$h$_2$, and
f$_1$h$_3$, respectively. Thus, of the three possible intermediate
configurations, only f$_1$h$_2$ is probable on a timescale comparable to that
for adatom diffusion (but of course cannot be excluded on longer timescales).
We note that the {\em barrier} for the f$_1$f$_2 \rightarrow$ f$_1$h$_2$
process --- 0.35 eV according to our calculations --- is very close to the
f$_1$h$_2$ configurational energy; thus, the barrier for the f$_1$h$_2
\rightarrow$ f$_1$f$_2$ process is vanishingly small (an unsignificant 0.01
eV) so that f$_1$h$_2$ is very short-lived.

If diffusion proceeds via a sequence of single-atom displacements, then there
are at this point two possibilities: (i) Atom A may jump to either h$_1$ or
h$_4$. (ii) Atom B may jump to either f$_4$ or back to f$_2$. It is easy to
see that, in either case, the next jump (of either A or B) would lead to an
improbable high-energy state of type f$_1$h$_1$ or f$_1$h$_3$, which both lie
substantially above the equilibrium f$_1$f$_2$ state. We can therefore only
conclude that diffusion by successive single-atom jumps only brings about a
local motion of the dimer, which is essentially trapped in a potential well
(possessing multiple minima) out of which it cannot escape on a timescale
appropriate to adatom diffusion.

We investigate next the possibility that diffusion proceeds via the concerted
motion of the two atoms forming the dimer.

\subsection{Diffusion by concerted jumps}\label{concert}

In order to go from one equilibrium site (such as f$_1$f$_2$) to another in a
concerted manner, the dimer must go through a metastable near-neighbour
hcp-hcp configuration, as evident from Fig.\ \ref{surf}. As already noted
above, the near-neighbour hcp-hcp configuration lies 0.30 eV above the
equilibrium fcc-fcc state; this is comparable to the single-adatom diffusion
barrier, and therefore is a possible candidate for the dimer diffusion path.

As depicted in Fig.\ \ref{surf}, there exists three possibilities for such a
concerted jump, labeled `cj$_1$', `cj$_2$', and `cj$_3$'. The paths `cj$_2$'
and `cj$_3$' are equivalent by symmetry, but different from `cj$_1$': in the
transition state of the cj$_1$ path, the two atoms forming the dimer sit on
either side of a surface atom, which is not the case for cj$_2$ and cj$_3$.
This makes diffusion extremely difficult in the cj$_1$ direction; our
calculations predict, indeed, a barrier of approximately 0.8 eV. In contrast,
for the cj$_2$ (or equivalently cj$_3$) process, we obtain a value of 0.37
eV. This is only 0.07 eV above the hcp-hcp configuration energy. Thus, once
in this state, the dimer can easily jump to an adjacent equilibrium fcc-fcc
configuration (which can be either the initial one or a new one) via the
cj$_4$ or cj$_5$ process indicated in Fig.\ \ref{surf}.

The barrier for the cj$_2$ process, 0.37 eV, is very close to that for adatom
diffusion; though we have not calculated the corresponding prefactors (they
are found experimentally to differ by roughly an order of
magnitude),\cite{kyuno} it is therefore certainly the case that both
processes will contribute significantly to diffusion at temperatures of
interest. Kyuno {\em et al.},\cite{kyuno} indeed, find the diffusivity of
dimers at 150 K to be comparable to that of adatoms at 100 K, with a
difference in activation barriers of 0.11 eV in favor of adatoms, larger than
that observed here (0.04 eV). Following Bogicevic {\em et
al.},\cite{bogicevic2} the temperature at which an Arrhenius process becomes
active can be estimated from $T_0=(E_A/k_B)/\ln(\nu_0/\Gamma)$ where $E_A$ is
the activation energy, $\nu_0$ is the attempt-to-diffuse frequency (prefactor),
and $\Gamma$ is the actual frequency at which diffusion is taking place. For
dimers, $E_A=0.37$ eV and $\nu_0 \approx 1.6 \times 10^{12}$ s$^{-1}$ (using
the prefactor determined experimentally);\cite{kyuno} with $\Gamma \approx 1$
s$^{-1}$, corresponding to an experimental deposition rate of 0.001--0.1
monolayer per second, one finds that dimers become active at approximately
150 K.

It is important to note that if only cj$_2$-type processes are possible, then
(because the barrier for cj$_1$-type processes is comparatively much larger)
diffusion would be constrained to a one-dimensional corridor, consisting of a
sequence of jumps such as cj$_2$-cj$_4$-cj$_2$-cj$_4$... Correlated jumps
need not, however, be exclusive of other processes. In section \ref{success},
we argued that successive jumps {\em alone} would not lead to mass transport,
which does not mean that they do not contribute. In fact, once in a hcp-hcp
configuration, the dimer can find its way to another corridor by the
combination of single-atom moves h$_1$h$_2 \rightarrow$ f$_1$h$_2
\rightarrow$ f$_1$f$_2$. The energies of the h$_1$h$_2$ and f$_1$h$_2$
configurations are comparable (0.30 vs 0.34 eV) and the barrier for the
f$_1$h$_2 \rightarrow$ f$_1$f$_2$ process is a smallish 0.01 eV; this
diffusion path is therefore highly probable. The corresponding barrier for
the concerted h$_1$h$_2 \rightarrow$ f$_1$f$_2$ process is, we have just
seen, 0.07 eV.

The error bar on the above values is of a few hundreths of an eV and we
therefore cannot determine precisely which route will be the preferred one.
It can however be safely concluded that either will lead to significant mass
transport: the limiting factor for diffusion is the f$_1$f$_2 \rightarrow$
h$_1$h$_2$ barrier of 0.37 eV, quite comparable to that for single atoms.

\section{Concluding Remarks}\label{concl}

Direct measurements of the diffusion of Pt atoms and dimers on Pt(111) using
low-temperature field-ion microscopy have been reported very recently by
Kyuno {\em et al.}\cite{kyuno} The activation energy for adatoms is found to
be $0.260 \pm 0.003$ eV while it is $0.37 \pm 0.02$ eV for dimers. The
barrier for adatoms is in excellent agreement with previous experimental
estimates,\cite{bott,feibelman1} which seems to rule out the possibility that
the activation barrier for adatoms as deduced from growth experiments is
``contaminated'' by contributions from dimers.

For dimers, the measured barrier is in remarkable agreement with our
theoretical estimate. Such a close agreement is probably to some extent
fortuitous as our calculations are precise to no more than a few hundreths of
an eV because of our neglect of dynamical and quantum effects, finite-size
limitations, and the approximate character of the LDA. Nevertheless, the
present study does establish that dimers are mobile at temperatures where
single-atom diffusion is active and can therefore contribute to mass
transport, albeit perhaps not in a very significant manner compared to
adatoms. Our calculations, further, indicate that the pathway for dimer
diffusion consists of a sequence of one-atom and concerted two-atom jumps.
Such a diffusion mechanism has been reported recently in the case of Al
dimers on Al(111) based on density-functional theory calculations (see Ref.\
cited in \onlinecite{bogicevic}).

\acknowledgements

We are grateful to Ari P Seitsonen and Normand Mousseau for useful
discussions. This work was supported by grants from the Natural Sciences and
Engineering Research Council (NSERC) of Canada and the ``Fonds pour la
formation de chercheurs et l'aide {\`a} la recherche'' (FCAR) of the Province
of Qu{\'e}bec. One of us (G.B.) is thankful to NSERC and FCAR for financial
support. We are grateful to the ``Services informatiques de l'Universit{\'e} de
Montr{\'e}al'' for generous allocations of computer resources. Part of the work
reported here has been performed on the IBM/SP-2 from CACPUS (``Centre
d'applications du calcul parall{\`e}le de l'Universit{\'e} de Sherbrooke'').


\narrowtext

\input{epsf.tex}

\vfill

\twocolumn
\narrowtext

\end{document}